\documentstyle[12pt,a4]{article}
\newcommand{\gsim}{\mbox{\raisebox{-1.0ex}{$\stackrel{\textstyle >}
{\textstyle \sim}$ }}}
\newcommand{\lsim}{\mbox{\raisebox{-1.0ex}{$\stackrel{\textstyle <}
{\textstyle \sim}$ }}}

\newcommand{\beq}{\begin{equation}}
\newcommand{\eeq}{\end{equation}}
\newcommand{\beqa}{\begin{eqnarray}}
\newcommand{\eeqa}{\end{eqnarray}}
\newcommand{\lmk}{\left(}
\newcommand{\rmk}{\right)}
\newcommand{\lnk}{\left\{ }
\newcommand{\rnk}{\right\} }
\newcommand{\lkk}{\left[}
\newcommand{\rkk}{\right]}
\newcommand{\llangle}{\left\langle}
\newcommand{\rrangle}{\right\rangle}
\newcommand{\bfx}{{\bf x}}
\newcommand{\bfX}{{\bf x}}
\newcommand{\bxt}{{\bf x},t}

\newcommand{\bfk}{{\bf k}}
\newcommand{\bfq}{{\bf q}}
\newcommand{\bfp}{{\bf p}}

\newcommand{\dirac}{\delta_{\rm drc}}
\newcommand{\etaone}{\eta_1}
\newcommand{\etatwo}{\eta_2}
\newcommand{\etathree}{\eta_3}
\newcommand{\etai}{\eta_i}
\newcommand{\zetaoo}{\zeta_{11}}
\newcommand{\zetatt}{\zeta_{22}}
\newcommand{\zetaot}{\zeta_{12}}
\newcommand{\zetaij}{\zeta_{ij}}

\newcommand{\ft}{\tilde{f}}

\newcommand{\lab}{\label}
\newcommand{\calG}{{\cal G}}
\newcommand{\calM}{{\cal M}}
\newcommand{\calN}{{\cal N}}
\newcommand{\calV}{{\cal V}}

\begin{document}
\baselineskip 10mm
\thispagestyle{empty}
{\baselineskip0pt
\leftline{\large\baselineskip16pt\sl\vbox to0pt{\hbox{\it Department of Physics}
               \hbox{\it Kyoto University}\vss}}
\rightline{\large\baselineskip16pt\rm\vbox to20pt{\hbox{KUNS-1416 }
           \hbox{YITP-96-44}
           \hbox{RESCEU No.39/96}
           \hbox{UTAP-243/96}
           \hbox{Oct 1996}
\vss}}%
}
\vskip1cm
\vskip1cm

\vskip1cm
\begin{center}{\large \bf NONLINEAR EVOLUTION OF THE GENUS STATISTICS
}

{\large \bf WITH ZEL'DOVICH APPROXIMATION
}

\end{center}
\vskip1cm

\begin{center}
 {\large
Naoki Seto} \\
{\em Department of Physics,~Kyoto University,~Kyoto 606-01,~Japan}
\end{center}

\begin{center}
 {\large Jun'ichi Yokoyama } \\
{\em Yukawa Institute for Theoretical Physics,~Kyoto University,~Kyoto 606-01,~Japan}
\end{center}

\begin{center}
 {\large Takahiko Matsubara}\\
{\em Department of Physics, School of Sciences and ~Research Center
for the Early Universe,\\ The University of
Tokyo, Tokyo 113, Japan
}
\end{center}
\begin{center}
 {\large Masaru Siino}\\
{\em Department of Physics,~Kyoto University,~Kyoto 606-01,~Japan}
\end{center}

\abstract{Evolution of genus density is calculated from Gaussian
initial conditions using  Zel'dovich approximation.
A new approach is introduced which formulates the desired quantity in
a rotationally invariant manner.  It is shown that normalized genus
density does not depend on the initial spectral shape but is a
function of the fluctuation amplitude only.  
}



\newpage
\section{INTRODUCTION} \label{sec:intro}

In the near future, we will obtain enormous information of the large-scale
structure (LSS) of the Universe  from several red-shift surveys. The ESO Slice
Project (Vettolani et al.\ 1995) and the Las Campanas Red-Shift Survey
(Shectman et al.\ 1995), have  recently been
completed, and data analysis of them are now being carried out
energetically.
 The Sloan Digital Sky Survey (SDSS)(Loveday 1996) to be completed in
the beginning of the next century
will provide us with a three dimensional map of the Universe with 
about one million galaxies, which means that the data available will be
one hundred times bigger  than those present.

Before the arrival of these  data, 
it is important to furnish statistical measures to analyze them so
that we will be able to probe the seed of density fluctuations as well 
as their evolution.
A number of measures have been proposed to characterize
statistical properties of LSS.  The most
commonly used quantity is the two-point correlation function
(Totsuji \& Kihara 1969), with  which we can find the amplitude of 
fluctuations on various scales but not their statistical distributions.
The counts-in-cells statistic, including the void
probability (White 1979), directly measures the statistical distribution of
galaxies and various theoretical models based on physical or
mathematical arguments have been proposed to fit the observational
data.  Although this measure contains mathematically full information
of the statistics in principle, it is difficult to relate the count
analysis with the visual image or connectivity of galaxy clustering
such as filamentary networks, sheet-like or bubble-like structures,
etc..
As a statistical measure to characterize such a topological structure
of galaxy distribution, the genus number has been used in the
analysis of recent redshift surveys (Gott, Melott, \& Dickinson 1986;
Gott, Weinberg, \& Melott 1987; Weinberg, Gott, \& Melott 1987; Melott,
Weinberg, \& Gott 1988; Gott et al. 1989; Park \& Gott 1991; Park, Gott,
\& da Costa 1992; Weinberg \& Cole 1992; Moore et al. 1992; Vogeley,
Park, Geller, Huchra, \& Gott 1994; Rhoads, Gott \& Postman 1994).

Theoretically, however, the value of the genus had been calculated
only  for the random Gaussian field (Adler 1981; Doroshkevich
1970; Bardeen  Bond, Kaiser, \& Szalay 1986 
(hereafter BBKS); Hamilton et al.~1986) 
for a long time
except for some restricted cases (Hamilton 1988, Okun 1990). 
Although the random Gaussian distribution is commonly assumed
as an initial condition,  nonlinear 
gravitational evolution of fluctuations certainly changes the
statistical distribution, and hence the Gaussian formula is not useful 
to trace  evolution.
The situation was somewhat improved recently as lowest-order
correction to the Gaussian genus number was analytically obtained by
one of us using the multi-dimensional Edgeworth expansion around the
Gaussian distribution (Matsubara 1994).  
Detailed comparison has also been done with
the results of $N$-body simulations, and it has been shown that the new
formula fits the numerical data well in the semi-linear regime but not
in the nonlinear regime (Matsubara \& Suto 1996).  This is in accord
with the fact that the one-point probability distribution function
(PDF) based on the Edgeworth series no longer fits the counts-in-cells
when the root-mean-square (RMS) value of the density contrast becomes
as large as $\simeq 1/4$ (Juszkiewicz et al. 1995; Ueda \& Yokoyama
1995).  
Thus it was
desired to find an analytic expression of the genus number for
realistic non-Gaussian distributions just as various non-Gaussian
models have been proposed for counts-in-cells statistics.

This was achieved by Matsubara \& Yokoyama (1996, hereafter MY) 
who calculated the
genus density in the case density field is characterized by a number
of Gaussian variables such as the lognormal  and the
chi-square distributions.  They have found that the lognormal formula 
fits evolution of a low-density cold-dark-matter (CDM) universe with
a cosmological constant well.  Although its agreement is excellent and 
the lognormal distribution is well motivated in the sense that it can
be obtained from the continuity equation in the nonlinear regime but
with linear or Gaussian velocity fluctuations (Coles \& Jones 1991), 
one cannot explain why it fits some models but not others, since it is
merely a fitting formula.

In the present paper we calculate nonlinear evolution of genus statistics 
 using the Zel'dovich (1970)  approximation (ZA) in order to clarify the
dependence of the results on the initial spectral shape of Gaussian
fluctuations.
 ZA describes  time evolution of density field more accurately
than the Eulerian perturbation theory in the semi-nonlinear regime. 
In particular, it has been claimed that ZA or its truncated version fit 
the pattern of the particles in $N$-body simulations well (Coles et
al.\  1993).

We make use of the Lagrangian description of ZA, in which
time evolution of density field is easily described using
initial fluctuations of gravitational potential, as well as the topologically
 invariant nature of the genus
number which does not depend on coordinate choice. 
For further  calculative simplicity, we introduce a
new idea to deal with the time evolution of the genus number density.
That is, we focus  on the events of  topology change of isovalue
surfaces in 3+1-dimensional spacetime and relate it with the desired
quantity with the help of
the Poincar\'{e}-Hopf theorem in differential topology.
Thus we can formulate the genus density in a rotationally invariant
manner and explicit evaluation becomes possible even if we must deal
with 31 Gaussian variables in ZA as explained below. 

The plan of the present article is as follows.
In \S 2 the basic concepts of genus statistics are  reviewed and we give an
expression when the density contrast is 
random Gaussian-distributed.
In \S 3 a brief description of ZA is given.
Then an important relation between the Lagrangian mapping and 
genus number is explained.
In \S 4 we discuss a new formulation of time evolution of genus number 
density. We can easily understand that this formalism is a natural
extension
of a well known peak formalism in a three dimensional case.
 Numerical results are described in \S 5.
\S 6 is devoted to conclusion and discussions.
We add four appendices. In Appendix A, we present a mathematical
foundation of our new formalism.  More intuitive discussion is given
in Appendix B.
In Appendix C, calculative procedure is explained in detail.
Finally in Appendix D, various perturbative formulae are given for comparison.

\section{GENUS CURVE IN THE GAUSSIAN DISTRIBUTION} \label{sec:gauss}
Genus is a topological quantity  defined by the number of the
homotopy classes of closed curves
that may be drawn on a surface without cutting it into two pieces.
The surface we consider in this article is the isodensity contour of
matter density field. We deal with the
comoving number density of the genus on these surfaces.
According to Morse theory there are a number of equivalent expressions of the
genus density,

\beqa
[{\rm genus~ density}]&=&
\rm {\frac{\lkk \# (holes)- \# (isolated~ regions) \rkk}{volume}}\\
 \nonumber \\
&=&{\rm {\frac {-\lkk \#(maxima)+ \#(minima)- \#(saddle~ point)\rkk}
{2 \ volume }}}
\eeqa
For example, in the case of  a 2-sphere, number of hole is 0, number of
isolated region is 1, and genus number is $-1$ using formula (1).
It has one maximum and one minimum along
any direction, so genus number calculated from expression (2) is also $-1$.
The genus number of a torus is 0, which we can easily confirm by
the both  formulae.
Usually we use the second formula, and evaluate the number
density of the stationary points to the
direction of the third spatial axis, $x_3$, together with the sign of the
determinant of the Hesse matrix to assign the proper signature to each 
stationary point depending on whether it is an extremum or 
a saddle.
Thus the genus number density, $G(\nu,t)$, of the contour  
 $\delta(\bfx,t)=\nu\sigma(t)$, 
where $\delta$ is density contrast  and $\sigma$ is its RMS,
can be expressed as follows (Doroshkevich 1970; Adler 1981; BBKS 1986).
\beq
G(\nu,t)=-\frac{1}{2}
\llangle \dirac\lmk \delta(\bxt)-\nu\sigma(t)\rmk\dirac(\etaone)
\dirac(\etatwo)|\etathree|(\zetaoo\zetatt-\zetaot^2)\rrangle,
\label{genusdef}
\eeq
where $\etai(\bxt)\equiv\partial_{x_i}\delta(\bfx,t)$,
$\zetaij(\bxt)\equiv\partial_{x_i}\partial_{x_j}\delta(\bfx,t)$, 
and $\dirac(\cdot)$ is Dirac's delta function.
Clearly  this formula (\ref{genusdef}) 
lacks $O(3)$ symmetry  due to the special treatment
of the third axis. We will overcome  this point later.
In the language of differential topology, the Morse vector
 field is identified with the vector field 
$(\partial_{x_1} x_3,~\partial_{x_2} x_3)$ in the above expression
(\ref{genusdef}),
where $x_3$ is regarded as a function of $x_1$ and $x_2$ implicitly
defined by
$\delta(x_1,~x_2,~x_3(x_1,~x_2);t)=\nu\sigma(t)$. (See Appendix
A for the definition of mathematical terminology.)

If $\delta$ is isotropic random Gaussian, then
we have  simple relations between the variance of 
$\delta,~~\eta$, and $\zeta.$
(BBKS 1986).
\beqa
 \langle\delta^2(\bxt)\rangle &\equiv& \sigma^2(t),
 ~~~\langle\delta(\bxt)\eta_i(\bxt)\rangle=0,~~~
 \langle\delta(\bxt)\zeta_{ij}(\bxt)\rangle=-\frac{\sigma_1^2(t)}{3}\delta_{ij}
 \nonumber\\
 \langle\eta_i(\bxt)\eta_j(\bxt)\rangle&=&\frac{\sigma_1^2(t)}{3}\delta_{ij},
 ~~~\langle\eta_i(\bxt)\zeta_{jk}(\bxt)\rangle=0,\\
 \langle\zeta_{ij}(\bxt)\zeta_{kl}(\bxt)\rangle&=&
 \frac{\sigma_2^2(t)}{15}(\delta_{ij}\delta_{kl}+\delta_{ik}\delta_{jl}
 +\delta_{il}\delta_{jk}),\nonumber
\eeqa
where $\sigma_1^2(t)$ and $\sigma_2^2(t)$ are defined,
respectively, by
\beq
  \label{eq:sigma}
  \sigma_1^2(t)\equiv \langle[\nabla\delta(\bxt)]^2\rangle,~~~{\rm and}~~~
  \sigma_2^2(t)\equiv \langle[\nabla^2\delta(\bxt)]^2\rangle.
\eeq
Then we can easily calculate the statistical average in equation
(\ref{genusdef}) to find
 (Doroshkevich 1970; Adler 1981; Gott, Melott, \& Dickinson 1986)
\beq
 G(\nu,t)=\frac{e^{-\frac{\nu^2}{2}}}{(2\pi)^2}
\lmk\frac{\sigma^2_1(t)}{3\sigma^2(t)}\rmk^{\frac{3}{2}}
 (1-\nu^2). \lab{lin}
\eeq
We should notice that $\sigma_2^2$ has no influence on equation (\ref {lin}).
This is because we only need sign information of the second derivative.

MY studied the genus number density in the case matter density field 
departs from random Gaussian.
They first evaluated genus curve in the lognormal density
distribution, whose statistical property is
characterized by a monotonic function, $i.e.$ logarithm,  
of a random Gaussian variable.
They also extended this kind of analysis to the
case density field is characterized by a number of 
Gaussian fields through a function as in the chi-square distribution. 
Their method is also applicable to the present case of ZA in which density
field is essentially characterized by six independent random Gaussian
variables and their spatial derivatives as explained in the next section.
 
\section{ZEL'DOVICH APPROXIMATION} \label{sec:zel}
\subsection{Basic Properties of ZA}
Zel'dovich Approximation is known as 
one of the most reliable approximation
to describe the weakly nonlinear evolution of density fluctuation 
(Shandarin \& Zel'dovich 1989).
It is a Lagrangian theory  giving the shift vector of
matter density field in terms of  
fluctuations in the initial gravitational potential, $\Phi_0$, as
 a function of the
 initial position $\bf{q}$.
The relation between Lagrangian coordinate, $\bf{q}$, and Eulerian coordinate,
$\bfx$, is given by
\beq
  \bfx = {\bf q}  -D(t)\nabla_{q} \Phi_0,   \label{xq}
\eeq
where $D(t)$ is a linear growth factor of small density fluctuations. 
The functional shape of $D(t)$ is 
determined by cosmological parameters. In this article $D$ is 
sometimes used as a time coordinate. 
ZA is confirmed to be a good approximation
to the stage when displacement $D(t)\nabla_{q} \Phi_0$ is not so
small, or until the   density fluctuations enter into a
nonlinear regime (Coles et al. 1993).

By the conservation of mass, the local density contrast in ZA,
$\delta_{ZA}$, is given  by the Jacobian of the
Lagrangian to Eulerian mapping.
\beqa
\delta_{ZA} (\bfx,t)+1
&=&\left|\frac{\partial (x_1,x_2,x_3)}{ \partial (q_1,q_2,q_3)}
\right|^{-1} \nonumber \\&
=&|(1-D \lambda _1)(1-D \lambda _2)(1-D \lambda _3)|^{-1} \nonumber \\&
=&\left|1-D{\rm tr} \alpha+\frac{D^2}{2} 
\{({\rm tr}\alpha)^2-{\rm tr}\alpha^2 \}-D^3 
{\rm det}\alpha\right|^{-1} \lab{dens}
\eeqa
where we have defined 
$\alpha_{ij} \equiv \partial_{q_i} \partial_{q_j}\Phi_0$, and 
$\lambda _i$'s are the three eigenvalues  of the tensor $\alpha_{ij}$.

Except for caustics, where density diverges and ZA breaks down,
the Lagrangian mapping is a homeomorphism. 
With the Lagrangian coordinate, ZA describes local density profile 
in a very simple manner. 
In order to calculate the genus number density of an isodensity
contour $\delta_{ZA}(\bfX,t)=\nu\sigma(t)$, it is convenient to
introduce a monotonic function of $\delta_{ZA}(\bfX,t)$ which may also depend
on time, $F[\delta_{ZA}(\bfX,t),t]$, and consider a contour
$F[\delta_{ZA},t]=F[\nu\sigma(t),t]$ to evaluate the genus number of
that surface which is clearly equivalent with the 
contour $\delta_{ZA}(\bfX,t)=\nu\sigma(t)$.  Choosing an appropriate
shape of $F$ greatly simplifies subsequent manipulations.
Specifically we choose $F$ as 
\beqa
  F[\delta_{ZA}(\bfX,t),t]&\equiv& \frac{1}{D(t)}\lmk 1-
  \frac{1}{1+\delta_{ZA}(\bfX,t)}\rmk \nonumber\\
  &=&\frac{1}{D}\lkk 1-
 \left|(1-D\lambda_1)(1-D\lambda_2)(1-D\lambda_3)\right| \rkk.
  \label{func}
\eeqa
Using (\ref{xq}) one can regard $F$ as a function of either $(\bxt)$
or $(\bfq, t)$, which is denoted by $F\equiv \ft(\bxt)$ and $F\equiv
f(\bfq,t)$, respectively.  Since we would like to obtain  genus
density in the Eulerian space we consider the contour $\ft(\bxt)=c$
below.
Following MY (eq.\ [18]) one can easily show that genus density of
this contour at $D(t)$, $\calG(c,D)$, is related with $G(\nu,t)$ by
\beq
  G(\nu,t)=\calG\lmk F[\nu\sigma(t),t],D(t)\rmk.   \label{relation}
\eeq
  
\subsection{Variables in ZA}
The fact that ZA contains more information, $i.e.$ three eigenvalues,
than the Eulerian perturbation theory makes it more reliable.  On the
other hand, calculation of genus number becomes much more difficult.
In fact we must deal with the following  variables.
\beqa
\alpha_{ij} &\equiv& \partial_{q_i} \partial_{q_j}\Phi_0,\\
\beta_{ijk} &\equiv& \partial_{q_i} \partial_{q_j}\partial_{q_k}\Phi_0,\\
\gamma_{ijkl} &\equiv& \partial_{q_i} \partial_{q_j}
\partial_{q_k}\partial_{q_l}\Phi_0.
\eeqa
We assume initial density fluctuations and so $\Phi_0$ are
random Gaussian-distributed.
Their correlation functions are then given by
\beqa
 \langle\alpha_{ij}\alpha_{kl}\rangle&=&8 {s_0}^2
(\delta_{ij}\delta_{kl}+\delta_{ik}\delta_{jl}+\delta_{il}\delta_{jk}),\\
&\equiv&{s_0}^2  d^{(4)}_{ijkl},\\
 \langle\alpha_{ij} \beta_{klm}\rangle&=&0,\\
 \langle\alpha_{ij}\gamma_{klmn}\rangle&=&-{s_1}^2  d^{(6)}_{ijklmn},\\
 \langle\beta_{ijk}\beta_{lmn}\rangle&=&{s_1}^2
d^{(6)}_{ijklmn},\\
\langle\beta_{ijk}\gamma_{lmno}\rangle&=&0,\\
 \langle\gamma_{ijkl}\gamma_{mnop}\rangle
 &=&{s_2}^2  d^{(8)}_{ijklmnop},
\eeqa
where $d^{(n)}_{\underbrace {ij\cdots m}_{n}}$ is an 
{\it n}-dimensional completely
symmetric tensor normalized as 
\beqa
d^{(n)}_{ii\cdots i}&=&n!.
\eeqa

In order to orthogonalize $\alpha$ and $\gamma$, we define new variables
$\omega_{ijkl}$ as
\beqa
\omega_{ijkl} \equiv \gamma_{ijkl}+\frac{3{s_1}^2}{2s_0^2} 
\alpha_{\{ij}\delta_{kl\}}-\frac{3{s_1}^2}{20 s_0^2} 
{\rm tr}\alpha d_{ijkl}^{(4)} \label{omega}
\eeqa
As can be easily checked  $ \langle\alpha_{ij} \omega_{klmn}\rangle$
vanishes and two body correlation of $\omega_{ijkl}$ is given by
\beq
\langle\omega_{ijkl}\omega_{mnop}\rangle=
d^{(8)}_{ijklmnop}-\frac{3{s_1}^4}{2s_0^2}
d^{(6)}_{mnop\{ij} \delta_{kl\}}+\frac{63{s_1}^2}{10 s_0^2}d^{(4)}_{ijkl} 
 d^{(4)}_{mnop}.  \label{omegacorr}
\eeq

Since $\alpha$, $\beta$, and $\omega$ are all Gaussian distributed,
we can express the probability distribution function (PDF), 
$P_L(\alpha ,\beta ,\omega)$, 
in terms of $s_0,~s_1$, and $s_2$ defined above. 
Denoting $\alpha,~\beta,$ and $\omega$ correctively by $y_1$,
$\cdots$, $y_{31}$, where six $\alpha$'s, ten $\beta$'s, and fifteen
$\omega$'s make up 31 variables in total, the PDF reads
\beqa
P_L(y_1, \cdots, y_{31})dy_1 \cdots dy_{31}&=& \frac{e^{-Q}}{[(2\pi)^n
{\rm det}\calN]^{1/2}}
dy_1 \cdots dy_{31} \lab{gauss},\\
Q&\equiv&\frac{1}{2}\sum_{i,j}^{} y_i {(\calN^{-1})_{ij}} y_j,\\
\calN_{ij}&\equiv&\langle y_i y_j\rangle.
\eeqa
Note that this is a PDF in the Lagrangian space.  We must calculate
the average using the Eulerian PDF, $P_E(y_1, \cdots, y_{31})$, which
would give a volume fraction in the {\it Eulerian} space in a ergodic system.
$P_E$ and $P_L$ are related by
\beq
  P_E(y_1, \cdots, y_{31})dy_1 \cdots dy_{31} =
  \frac{J(\bfx;\bfq)}{\langle J(\bfx;\bfq)\rangle}
  P_L(y_1, \cdots, y_{31})dy_1 \cdots dy_{31},  \label{euler}
\eeq 
where $J(\bfx;\bfq)\equiv \partial(x_1,x_2,x_3)/\partial(q_1,q_2,q_3)$
is the Jacobian and $\langle J(\bfx;\bfq)\rangle$ is its average with
respect to the Lagrangian PDF $P_L$ (Kofman 1994).
Since only the combination of  $D s_0$ has a physical meaning, we may
set $s_0=1$ to fix normalization of $D$.  However, we will put $s_0$
explicitly when appropriate below.
Bernardeau \& Kofman (1995) derived an analytic formula of one-point
density PDF using ZA. 
Their result is rather complicated due to the divergence
of density fields at the caustics points.  Furthermore
RMS amplitude of fluctuation diverges in ZA.  Hence we will adopt
the variance in linear theory, $\sigma_l(t) \equiv \sqrt{120}s_0D(t)$, 
to normalize the amplitude of $\delta$.

Since $\ft$ reduces to a linear combination of
$\alpha$ in the limit  $D\longrightarrow 0$, 
$\ft(\bfx,0)={\rm tr}\alpha$, it is
also random Gaussian then.
Therefore at $D=0$, when the Eulerian coordinate coincides with the
Lagrangian counterpart, the genus density, $\calG(c,0)$ is easily
obtained in the form given in {\S} 2,  
\beqa
{\cal G}(c,0) &=& 
\frac{14^{\frac{3}{2}}}{\left(2 \pi \right)^2}  \frac {{s_1}^3}{ {s_0}^3} 
\exp \left(- \frac{ c^2}{240s_0^2}\right)\lmk 1-
\frac{c^2}{120s_0^2}\rmk.
\eeqa

When $D$ is finite, $\ft(\bfx,D)$ is now a nonlinear combination of
$\alpha$ and 
calculation would become extremely complicated and tedious if we used
the usual expression (\ref{genusdef}) without rotational 
symmetry due to the fact that we tried to count maximums, minimums,
and saddle points of an isodensity surface along the $x_3$-direction.
In the next section we
introduce a new approach which greatly simplifies evaluation.

\section{3+1 FORMALISM}
\label{sec:3+1}

\subsection{Basic Idea}

Here we  present a 
new formulation based on the fact that time evolution of genus number
density can be expressed in an $O(3)$ invariant manner.
We know genus number density $\calG(c,0)$ on the three dimensional
spacelike hypersurface 
$\Sigma_0$ at $t \to 0$ or $D(t) \to 0$ as in \S 2.  We can trace its time 
evolution making use of the fact that
the points (events) where genus number changes ( {\bf P}, {\bf Q}, and
{\bf R}, in 
Figure 1) correspond to  a minimum, a maximum,
or a saddle point, namely, 
stationary points of an isovalue hypersurface of $\ft(\bfx,D)=c$ 
along the time- or $D$-direction in
four dimensional spacetime ($x_1$, $x_2$,  $x_3$, $D$).
The number density of these  stationary points  can be 
statistically expressed as in three  
dimensional case following  similar argument to derive
equation (\ref{genusdef}). 
To evaluate the statistical average of
 the frequency of these events is essential in our new
formalism. This idea is 
mathematically supported by a relationship derived from
Poincar\'{e}-Hopf 
theorem, an important theorem in differential topology (see Appendix A):
\beq
2 \cdot {  ind(X,~\calV)}=\chi (\Sigma_1)-\chi (\Sigma_0).
\eeq
Here $\calV$ is a three dimensional manifold which has two two-dimensional
boundaries, $\Sigma_0$ and $\Sigma_1$.
In our case $\calV$ is a three dimensional isodensity contour embedded 
into the four dimensional spacetime, and $\Sigma_0$ and $\Sigma_1$ are 
the initial and the final
two-dimensional spatial section of  constant time or $D$,
respectively.
$ ind(X,~\calV)$ is index of Morse vector field $X$ in $\calV$, 
and $\chi (\Sigma_i)$
is Euler number in each spatial section $\Sigma_i$, which 
is related to genus number $g$ as $\chi=-2 g$.
  This formula shows that  genus number in the final  surface  $\Sigma_1$ 
of interest is determined  by that of initial surface $\Sigma_0$
and the information of stationary points in three dimensional region $\calV$.

In the present case, the Morse vector field $X$ is taken as,
\beq
X=(\partial_{x_1} D_c,~\partial_{x_2} D_c,~\partial_{x_3} D_c),
\eeq
 where $D_c$ is a function of $(x_1,~x_2,~x_3)$ 
implicitly defined by
\beq
\ft (x_1,~x_2,~x_3,~D_c(x_1,~x_2,~x_3))=c. \label{dcdef}
\eeq
As explained in Appendix B, we can classify stationary points into four
groups by the signature of the Hesse Matrix 
${\cal H}_{ij}\equiv \partial^2 D_c/ \partial
x_i \partial x_j$.
\begin{center}
 \begin{tabular}{lcr}
type  &signature & $\Delta g$\\
\it 1 creation &$+ + +$&$-1$\\
\it 2 merging &$+ - -$&$-1$\\
\it 3 annihilation &$- - -$&$+1$\\
\it 4 split &$+ + -$&$+1$
\end{tabular}
\end{center}
$1 \leftrightarrow 3$ and $2 \leftrightarrow 4$ are related 
to each other by time reversal.
Here one should notice that the signatures 
and signs of the Hesse Matrix  are related to  changes of
genus numbers, $\Delta g$, as,
\beq
\prod({\rm signature})={\rm sgn(det(\cal H))}=-\Delta g.
\eeq
So we have only to evaluate four dimensional correspondence of  
equation (\ref{genusdef}).

After simple consideration, we find the net number 
density, $N_{st}$, of these stationary
points including sign information in  four dimensional space as
\beq
N_{st}=-
 \dirac \lmk \ft(\bfx,D)-c\rmk\dirac(\partial_{x_1}\ft)
\dirac(\partial_{x_2}\ft)\dirac(\partial_{x_3}\ft)
\partial_D\ft {\rm det}(\partial_{x_i}\partial_{x_j}\ft). \label{stationary}
\eeq
This is a natural extension  of (\ref{genusdef}) to a three dimensional
hypersurface. 
So the genus density ${\cal G}$ on the three 
dimensional space $\Sigma_{1}$ on which $D=D_1$ is given by 
\beq
{\cal G}(\ft =c,~D_1)={\cal G}(\ft=c,~0)+ 
\int_{0}^{D_1} \llangle N_{st} \rrangle _{\alpha \beta \omega}dD,
  \label{total}
\eeq
where the average $\langle\cdots\rangle_{\alpha \beta \omega}$ is to
be calculated with the Eulerian PDF (eq.\ [\ref{euler}]).

\subsection{Method for Calculation of  $ \llangle N_{st} \rrangle$}

In order to calculate $\llangle N_{st} \rrangle$ we first rewrite equation
(\ref{stationary}) in terms of
$f(\bfq,D)=\ft(\bfq -D\nabla_\bfq\Phi_0,D)$.
From
\beqa
\frac{\partial\ft}{\partial x_1}&=&
f_{,l}\frac{\partial q_l}{\partial x_i}, \nonumber\\
\frac{\partial^2 \ft}{\partial x_i\partial x_j} &=&
f_{,lm}\frac{\partial q_l}{\partial x_i}\frac{\partial q_m}{\partial x_j}
+f_{,l}\frac{\partial^2 q_l}{\partial x_i\partial x_j}, \\
\frac{\partial\ft}{\partial D}&=&\frac{\partial f}{\partial D}  
+ \frac{\partial\Phi_0}{\partial q_l}\frac{\partial q_m}{\partial x_l}f_{,m},
\nonumber
\eeqa
where $f_{,l}\equiv \partial f/\partial q_l$ etc., 
we find
\beq
N_{st}=-
 \dirac \lmk f(\bfq,D)-c\rmk\dirac(f_{,1})
\dirac(f_{,2})\dirac(f_{,3})
f_{,D} {\rm det}(f_{,ij})|J(\bfx;\bfq)|^{-1}. 
\eeq
Hence we obtain
\beqa
\llangle N_{st} \rrangle &=&\int N_{st}P_E(y)d^{31}y \nonumber\\ 
&=& \frac{-\int \dirac \lmk f(\bfq,D)-c\rmk\dirac(f_{,1})
\dirac(f_{,2})\dirac(f_{,3})
f_{,D} {\rm det}(f_{,ij})P_L(y)d^{31}y}{\langle J(\bfx;\bfq)\rangle}.
\eeqa
The above equality can be interpreted as follows.  The number of
stationary points in a certain region remains invariant whichever
coordinate, Lagrangian or Eulerian, one uses.  So its density changes
only by the volume factor represented by the Jacobian
$\langle J(\bfx;\bfq)\rangle$, which has been calculated by Kofman et
al. (1994).

In the above formulation 32-dimensional integration must be accomplished
in principle including $D$-integral.  Making full use of the
rotational symmetry, however, we can perform most of these integrals
such as those over $\beta$ and $\omega$ algebraically following the
scheme introduced by MY.

First we note that thanks to the $O(3)$ invariance  $ N_{st} $ does
not depend on the Euler angles of the transformation
to principal axes of $\alpha_{ij}$, and we may assume that
$\alpha_{ij}$ is diagonalized from the beginning without loss of generality.
So among six integrals over $\alpha_{ij}$, those corresponding to the 
angular integrals are trivial (see Appendix B of BBKS).
Consequently we can replace 
\beq
\alpha_{11}=\lambda_1,~~ \alpha_{22}=\lambda_2,~~
\alpha_{33}=\lambda_3,~~
\alpha_{12}=\alpha_{23}=\alpha_{31}=0.  \label{diagonal}
\eeq

We now turn to averaging over $ \beta$ and $\omega$ whose details are
given in Appendix C.
The determinant of $f_{,ij}$  in $ N_{st} $ has $3!=6$ terms.
Thanks to the $O(3)$ symmetry, we may identify
$f_{,11}f_{,23}f_{,32},~ f_{,22}f_{,13}f_{,31}$, and
$ f_{,33}f_{,12}f_{,21}$ with each other
in the expression to be averaged.
We can therefore rearrange  ${\rm{det}} f_{,ij}$ as follows.
\beq 
{\rm det} f_{,ij} \Rightarrow f_{,11}f_{,22}f_{,33}-3f_{,11}f_{,23}f_{,23}+
2f_{,12}f_{,23}f_{,31}.
\eeq
Furthermore, since the Gaussian average of odd multiple is equal to zero,   
we can drop odd terms of $\omega_{ijkl}$ or $\beta_{ijk}$. 
Using these facts it is not so hard to calculate the integrals.
 
First, integrate over $\omega_{ijkl}$. 
We only have to replace $\omega\omega$ by $\llangle
\omega\omega\rrangle $, 
terms containing
$\omega\omega\omega$ or $\omega$  by 0, 
and do not change the terms with no  $\omega$.
Next, integrate over $\beta_{ijk}$. Using three delta functions $\dirac(f_i)$,
we can independently eliminate integration over $\beta_{111}$,
$\beta_{222}$,  and $\beta_{333}$.
As the rest of  $\beta_{ijk}$-integral (seven dimensional) is ordinary 
Gauss integral, we can 
algebraically manage them by such  a formula
\beq
\llangle \beta_a \beta_b\beta_c\beta_d \rrangle=M^{}_{ab}M^{}_{cd}+
M^{}_{ac}M^{}_{bd}+M^{}_{ad}M^{}_{bc},
\eeq
where $M^{}_{ab}$ is the correlation matrix 
$\llangle \beta_a \beta_b \rrangle$.
This matrix is $7\times 7$ 
but constituted from 
$2\times 2  \oplus2\times 2  \oplus 2\times 2  \oplus1\times1$ blocks 
due to our  choice of diagonalized $\alpha$.

Finally we numerically integrate over $\lambda_i$ and $D$.
The PDF of $\lambda_1$, $\lambda_2$, and $\lambda_3$,
has been  given by Doroshkevich (1970).
\beqa
 P(\lambda_1,\lambda_2, \lambda_3) d\lambda_1 d\lambda_2 d\lambda_3
&=&\frac{5^{\frac{5}{2}} 27}{48 \pi {\sigma_0}^6}|( \lambda_1-\lambda_2)
( \lambda_2-\lambda_3)( \lambda_3-\lambda_1)| \nonumber\\
& & \times \exp\left[ -\frac{1}
{{\sigma_0}^2} \left( 3 {J_1}^2 -\frac{15}{2} {J_2}\right)\right]
d\lambda_1 d\lambda_2 d\lambda_3,
\eeqa
where we have defined
\beqa
J_1&\equiv&\lambda_1 +\lambda_2+\lambda_3,\\
J_2&\equiv&\lambda_1\lambda_2+\lambda_2\lambda_3+\lambda_3\lambda_1,\\
\sigma_0&\equiv& \sqrt{120} s_0.
\eeqa
There is a subtle point in the integration over $\lambda_i$, which is
the emergence  of caustics where $1-D\lambda_i$ vanishes.
Since these points make both ZA and our homeomorphic picture 
break down, our result would not be trustworthy if the region 
$1-D\lambda_i\le 0$ should affect the final result. 
For this purpose we integrate over $\lambda_i$ with two different
ways, one over the full range of $\lambda_i$ and the other over the region 
$1-D\lambda_i\ge 0$ only. These two results should agree with each
other for our analysis to be justified.  Note that even though 
our homeomorphic picture breaks down at caustics we do not encounter
any divergent quantities in our analysis there unlike the case of
one-point PDF studied by Bernardeau \& Kofman (1995).

Because of $\dirac(f-c)$  in equation (\ref{diagonal}), 
we can eliminate $\lambda_1$. So the number of 
 dimensions to integrate is only three, $\lambda_2$, $\lambda_3$, and $D$.
The function to be  integrated is complicated very much,
whose abbreviated form is given at the end of 
Appendix C, but 
it is easily  dealt with  numerically
for the rapid convergence due to the overall 
Gaussian-like factor from the PDF of $\lambda$.

\section{RESULTS} \label{sec:}
Before presenting the numerical results we comment on the functional
dependence of the final results of 
\beq
{\cal G}(f=c,~D_1)={\cal G}(f=c,~0)+ \int_{0}^{D_1} \llangle N_{st} \rrangle 
_{\alpha \beta \omega}dD \lab{45}.
\eeq
As pointed out before, equation (\ref{45}) does not contain  $s_2$. This is because
we only need sign information of second derivative and not their
amplitude.
Furthermore in ZA  case,  $s_1$ appears in equation (\ref{45}) only as an
overall factor ${s_1}^3/{s_0}^3$ as in the Gaussian formula 
(eq.\ [\ref{lin}]).
This implies an interesting fact that 
for any value of    $s_1$,
every genus curve has a similar shape parameterized only by  
$D$ since we have set $s_0=1$.
Therefore the ratio $ {\cal G}(c,~D_1)/{\cal G}(0,~D_1)$ 
depends on $D$ and $c$ only and not even on $s_1$.
 From equation (\ref{relation}) genus density of the surface
$\delta=\nu\sigma_l(t_1)$ is given by
\beq
  G(\nu,D_1)=\calG\lmk\frac{\nu\sigma_l}{D_1(1+\nu\sigma_l)},D_1\rmk.
\eeq

We now proceed to the numerical results carrying
 out numerical integration with different final time ({\it i.e.}, 
$\Sigma_{D_1}$).
We can arbitrarily chose the normalization of  $s_0$ as pointed out before.
On the other hand  $s_1$ changes only the scale of vertical
coordinate for the overall factor.
Since we concentrate on the ratio
$G(\nu,~D_1)/G(0,~D_1)$  
we can also put  $s_1=1$ in the calculative process.

In Figures 2(a)-(d), we plot the genus curves at $\sigma_l=0.1,
~0.22, ~~0.5,$ and 1.0.  
For comparison, we have also plotted the results of two perturbative
formulae, 
second-order Eulerian perturbation  (eq.\ [\ref {eq:eds2nd}]) and 
ZA to the same order (eq.\ [\ref {eq:zel2nd}]).
These curves show that all of them are very close to each other 
as expected in the 
region where density contrast is small ($\sigma_l|\nu| \lsim 1/4$), but 
deviate considerably in both ends ($\sigma_l|\nu| \gsim 1/4$).
In the same figures we have also depicted  contribution of
the first term in equation (\ref{45}) to show the importance of time
evolution of stationary points.

In Figure 2(d) with $\sigma_l=1$ we have depicted two curves of ZA
result.  The solid line corresponds to the case integration range of
$\lambda_i$ has been restricted to the region $1-D\lambda_i \geq 0$,
while short-dash-dotted line represents the case integration has been
done over the full range of $\lambda_i$.  Their discrepancy is at most 
about $10\%$ indicating the accuracy of our results.  For the cases
with smaller $\sigma_l$ these two curves practically coincide
 with each other so only one of them has been depicted in Figures 2(a)-(c).

Our results show clearly that the peak of the genus curve based on ZA
shifts to right
or to the region $\nu > 0$. The same tendency is observed in the result of 
Eulerian perturbation theory without smoothing (Matsubara 1994).
However, the peak of the genus curve shifts to left 
in the latter theory once uniform  smoothing 
such as a Gaussian smoothing with a specific width, say $8h^{-1}$Mpc,
is applied.
The results of CDM type $N$-body  
simulations also exhibits shift to  left after the same  smoothing.
Unfortunately we cannot apply 
the same sort of smoothing to the final configuration
in our result because we are working in the Lagrangian coordinates
which coincides with the Eulerian counterpart only at the initial
epoch and because smoothing and time evolution do not commute with
each other.

The information of initial condition must be contained more directly in 
Lagrangian description 
than in Eulerian description. So some of it might have been lost 
in Euler-type smoothing. The discrepancy of the shift of the peak 
might support this view implicitly.
Therefore a new,  more adequate  smoothing method is desired.
This issue is under investigation now and we do not go into it further 
here.

\section{CONCLUSION}

We studied the gravitational evolution of genus number density of
isodensity surface in terms of ZA. 
While ZA is more reliable than the Eulerian perturbation theory to
describe the density profile in semi-nonlinear regime, it is much more
difficult to calculate genus density in the former which depends on  
more variables.  We have overcome this difficulty by introducing a new 
approach in which we focus on evolution of genus number in the 3+1 dimensional
spacetime rather than dealing with that at a specific time from the
beginning.  We can formulate the former in a rotationally-invariant
manner so that its calculation is easier and relate it with the
desired quantity making use of the Poincar\'e-Hopf theorem in
differential topology.  Intuitively, this formula is based on the fact
that events of topology change of equal-time spatial isodensity
surfaces can be identified with stationary points of the corresponding 
hypersurfaces in four dimensional spacetime as explained in Appendix
B.
Without such a new idea we would have to have dealt with
31 dimensional numerical integration instead of just three.

We also mention that the Lagrangian nature of ZA is very suitable for
our formulation.  On the other hand, that made it impossible to apply
uniform smoothing to the final result and so we have been unable to
compare our result with observations or simulations.  In the
Lagrangian description information on the initial condition is
reflected more directly hence it is desirable to invent a new
smoothing method of discrete data which is compatible with Lagrangian
theories in order to find out the nature of primordial fluctuations.
We might obtain such a means by changing the smoothing width as a 
function of local density which is often adopted in smoothed particle
simulations.  However, we should clarify theoretical basis before
drawing final conclusion.  This issue is under investigation now.

Apart from problems of smoothing we have shown that normalized genus
density is a function of $D$ only and does not depend on the initial
spectral shape as long as initial fluctuation is Gaussian and 
ZA is applicable.

\vskip1cm

NS would like to thank Professor H.\ Sato for his continuous encouragements.
This work was partially supported by the Japanese Grant
in Aid for Science Research Fund of the Ministry of Education, Science,
Sports and Culture No.\ 08740202(JY) and No.\ 5391(MS).

\begin{center}
\large\bf{APPENDICES }
\end{center}

\appendix

\section{POINCAR\'{E}-HOPF THEOREM}

In this appendix we review a mathematical basis of our new formulation.
The following
Poincar\'{e}-Hopf theorem (Milnor 1965) is essential for our analysis.
\begin{description}
\item[\sl Theorem I] (Poincar\'{e}-Hopf)
        Let $\calM^n$ be a compact $n$-dimensional ($n\geq 2$) 
$C^{\infty}$ manifold. $X$ is 
        any $C^{\infty}$ vector field with at most a 
finite number of zeros, which are the points with vanishing 
magnitude of $X$, satisfying the following two conditions. 
\begin{description}
\item[(a)] The zeros of $X$ are contained in $Int \calM^n$. 
\item[(b)]$X$ has outward directions on $\partial \calM^n$.
\end{description} 
Then the sum of the indices of $X$ at all its zeros is 
equal to the Euler number $\chi$ of $\calM^n$,
        \begin{equation}
                        ind(X)=\chi(\calM^n).
        \end{equation}
\end{description}

\noindent
Here the index of the vector field $X$ at a zero, $P$, is defined as follows. 
Let $X_a(x)$ be the components of $X$ with respect to local coordinates 
$\{x^a\}$ in a neighborhood of $P$. Set $v_a(x)=X_a(x)/\vert X\vert$. If 
we evaluate $v$ on a small sphere centered at $x(P)$, we can regard 
$v_a(S^{n-1})$ as a $C^{\infty}$ mapping from $S^{n-1}$ into $S^{n-1}$. 
The mapping degree of this map is called the index of $X$ at the zero $P$.
For example, if the map is homeomorphic, the mapping degree of the 
orientation-preserving 
(reversing) map is $+1$ ($-1$). 

In discussing genus statistics we identify the vector field $X$ with a 
gradient of a $C^{\infty}$ function $h$ called the Morse function.  
If $P$ is a non-degenerate critical point of this function with Morse 
index $\lambda$, which is the number of 
the negative eigenvalues of Hesse matrix $\partial_a\partial_b h$, 
then the index of the gradient field $\partial_a h$ at $P$ is 
$(-1)^{\lambda}$.  
In the present paper $D_c(x_1,x_2,x_3)$ defined in equation (\ref{dcdef})
serves as the Morse function.

We treat three dimensional manifold embedded in 
a four dimensional spacetime manifold as an isovalue contour. The three 
dimensional manifold has two two-dimensional boundaries as an initial 
boundary and a final boundary. For such a manifold, we use the following 
modification of the Poincar\'{e}-Hopf theorem. 
Now we consider odd-dimensional manifold with two boundaries 
$\Sigma_{0,1}$. 

\begin{description}
        \item[\sl Theorem II](Sorkin 1986)
                Let $\calM^n$ be a compact $n$-dimensional 
($n> 2$ is an odd number) $C^{\infty}$ manifold 
                with $\Sigma_0\cup\Sigma_1=\partial \calM^n$ and 
                $\Sigma_0\cap\Sigma_1=\phi$. $X$ is 
        any $C^{\infty}$ vector field with at most a 
        finite number of zeros, satisfying the following two 
        conditions:  
\begin{description}
\item[(a)] The zeros of $X$ are contained 
        in $Int \calM^n$. 
\item[(b)] $X$ has inward directions at $\Sigma_0$ and outward 
        directions at $\Sigma_1$. 
\end{description}
        Then the sum of the indices of $X$ at all its zeros is related to 
the Euler number of $\Sigma_0$ and $\Sigma_1$ as
        \begin{equation}
                \chi(\Sigma_{1})-\chi(\Sigma_{0})=2 \cdot ind(X).
                \label{eqn:ss}
        \end{equation}
\end{description}
{\sl Proof:} 
\ Let $\tilde{\calM}^n$ be a manifold obtained by gluing 
$\calN=\Sigma_0\times[0,1]$ and $\calM^n$ at $\Sigma_0\times[1]$. Also let 
$\tilde X$ be an extension of $X$ to $\tilde{\calM}^n$ which has at most a 
finite number of zeros and is outward directed on 
$\Sigma_0\times[0]\subseteq \calN$. Since $\tilde X$ and its restriction 
$\tilde X_\calN$ to $\calN$ are outward directed on 
the entire boundaries of $\tilde{\calM}^n$ and $\calN$, 
respectively, we can apply the theorem I 
to them. Then we have
\begin{equation}
        ind(\tilde X)=\chi(\tilde{\calM}^n),
\end{equation}
and
\begin{equation}
        ind(\tilde X_\calN)=\chi(\calN).
\end{equation}
Since $\calM^n$ and $\tilde{\calM}^n$ are diffeomorphic, 
$\chi(\calM^n)=\chi(\tilde{\calM}^n)$. 
Using $\chi({\cal A}\times {\cal B})=\chi({\cal A})\chi({\cal B})$ and 
$ind(\tilde X)=ind(X)+ind(\tilde X_\calN)$, we obtain
\begin{equation}
        ind(X)=\chi(\calM^n)-\chi(\Sigma_0).
        \end{equation}
Similarly we find
\begin{equation}
        ind(-X)=\chi(\calM^n)-\chi(\Sigma_1),
        \end{equation}
by interchanging the roles of $\Sigma_0$ and $\Sigma_1$.
Since $ind(-X)=(-1)^n ind(X)$ (inversion of $S^{n-1}$ reverses orientation 
only if $n$ is odd), we have equality (\ref{eqn:ss}).

Finally we consider time-vector field $X=\nabla D_c$. As mentioned above, 
the index of the vector field $X$ is given by the number of negative 
eigenvalues of a corresponding Hesse matrix $\partial_a\partial_b D_c$. 
Therefore $ind(X)$ in theorem II is evaluated in the same manner as in
Doroshkevich (1970).

\section{STATIONARY POINTS IN 3+1 SPACETIME}

In this appendix we present an intuitive explanation  of the 
relation between time evolution of genus number of two dimensional
surfaces in three dimensional space and stationary points along the
time axis on three
dimensional hypersurfaces consisting of their trajectories in four
dimensional spacetime.
These points  correspond to events where
genus number of a two dimensional contour changes.
They are classified into four groups by the signature of the
Hesse matrix.
The change of the genus number turns out to be   minus of
the sign of the determinant of that matrix as explained below.

First we consider the case when a new isovalue 
surface is created at a spacetime
point  {\bf P} with time $t=t_P$, namely, 
when genus number changes by $-1$ as shown in 
equation (1).
We introduce a local coordinate with its origin at {\bf P} and
principal axes denoted by X, Y, and Z in Figure 3-1.  In the upper
figures emergence of a new contour of $f(X,Y,Z;D)=c$ or $D=D_c(X,Y,Z)$
 at $t_P$ is
depicted, showing time evolution from $t_P-\varepsilon$ to
$t_P+\varepsilon$ $(\varepsilon > 0).$  Lower-left figures represent
how intersecting points of the surface to each principal
axis move with time, which is summarized in the lower-right figure 
showing evolution of the contour along each axis. 
In the present coordinate system we have only to investigate the sign
of second derivatives ${\partial}^2 D_c / {\partial} \rm{X}^2 $,
 ${\partial}^2 D_c / {\partial} \rm{Y}^2$, and
 ${\partial}^2 D_c / {\partial} \rm{Z}^2$
of the  three dimensional contour $D=D_c(X,Y,Z)$
in four dimensional space.
Apparently all of these derivative is positive, so the  signature of the
Hesse matrix is $+++$, and its determinant  is positive, too.
Conversely when signature is $+++$ the type of the event is creation. 
Since annihilation is its time reversal  the signature is $---$.

Next we consider the case when two surfaces  merge at a spacetime
point {\bf Q} which is depicted in Figure 3-2 in which the origin is
now taken at {\bf Q}.  Depending on whether the two surfaces are connected or 
disconnected this event is interpreted as merging of two isolated
surfaces or creation of a hole.
>From the lower-right figure we can easily convince ourselves that 
the signature of Hesse matrix there is $++-$, 
and  genus number changes by $+1$.
Since split, which corresponds to either increase of an isolated region or
annihilation of a hole,  is time reversal of this, the signature is
$+--$ and genus changes by $+1$.

We can summarize the above results that the change of genus number at
each event is equal to minus the signature of the Hesse matrix.

\section {CALCULATION SCHEME}
In this appendix, we briefly explain the calculative procedure of 
$ \llangle N_{st} \rrangle _{ \beta \omega}$
(see eq.\ [\ref{stationary}]).
This is essentially an extension of the scheme developed by MY.
We perform $\omega$- and $\beta$-integral algebraically. 
The remaining integration over $\lambda$ and $D$ has to be done numerically.
Throughout this appendix we put $s_0 =1$. 

In the quantity to be averaged,
\beq
N_{st}=-
 \dirac \lmk f(\bfq ,D)-c\rmk\dirac(f_{,1})
\dirac(f_{,2})\dirac(f_{,3}) f_{,D} {\rm det}(f_{,ij}),
\eeq
$\omega$ appears only in the second derivative $f_{,ij}$ and $\beta$
appears both in $f_{,i}$ and $f_{,ij}$ as
\beqa
f_{,i}&=& \frac{\partial f}{\partial\alpha_{jk}}\beta_{ijk}, \\
f_{,ij}&=&  {\frac{\partial^2 f} {\partial \alpha_{kl}
 \partial \alpha_{mn} }} \beta_{ikl} \beta_{jmn}+
{\frac{\partial f}{\partial \alpha_{kl}}} \gamma_{ijkl}  \nonumber \\*
&=& {\frac{\partial^2 f} {\partial \alpha_{kl} \partial \alpha_{mn} }}
  \beta_{ikl} \beta_{jmn}+ \frac{\partial f}{\partial \alpha_{kl}}
\ \omega_{ijkl}  \nonumber \\*  
&-& \frac{\partial f}{\partial \alpha_{kl}} \lnk \frac{3}{2} {s_1}^2
\alpha_{\{ij }\delta_{kl\}} 
-\frac{6}{5} {s_1}^2
\lmk \alpha_{11}+\alpha_{22}+\alpha_{33} \rmk \lmk 
\delta_{ij}\delta_{kl} +\delta_{ik}\delta_{jl}
+\delta_{il}\delta_{jk}\rmk  \rnk.
\lab{c3}
\eeqa
Since $\alpha_{ij}$ has only diagonal components, 
$\alpha_{ij}=\lambda_i\delta_{ij}$, 
we find $g_{ij} \equiv \partial f / \partial \alpha_{ij}$ 
in a very simple form below.
\beqa
g_{11}&\equiv &  \frac{\partial f }{ \partial \alpha_{11}}=
 {\frac{\partial f }{ \partial \lambda_{1}}}=-(1-D \lambda_2)(1-D
   \lambda_3) \equiv g_1, \\
g_{22} &=& -(1-D \lambda_1)(1-D \lambda_3) \equiv g_2,\\
g_{33} &=& -(1-D \lambda_1)(1-D \lambda_2) \equiv g_3,\\
g_{ij}&=&0  ~~~~~(i \ne j).
\eeqa
Similarly
$h_{ijkl} \equiv \partial ^2 f / \partial \alpha_{ij}
   \partial \alpha_{kl}$ reads,
\beqa
h_{1122}&=&h_{2211}=-h_{1221}=-h_{2112}=D(1-D \lambda_3) \equiv h_3,\\
h_{1133}&=&h_{3311}=-h_{1331}=-h_{3113}=D(1-D \lambda_2) \equiv h_2,\\
h_{2233}&=&h_{3322}=-h_{2332}=-h_{3223}=D(1-D \lambda_1) \equiv h_1,\\
h_{ijkl}&=&0~~~~~({\rm others}). \nonumber
\eeqa

As explained in the text, using $O(3)$ symmetry the determinant of $f_{,ij}$
can be replaced by
\beq
{\rm det}f_{,ij} \Rightarrow
 f_{,11} f_{,22}  f_{,33} -3  f_{,11} f_{,23}^2 + 2f_{,12}f_{,32} f_{,13}. \lab{c11}
\eeq
For notational simplicity we rewrite  $\beta_{ijk}$ as follows.
\beqa
a_1 &\equiv& \beta_{221},~~~a_2 \equiv \beta_{331},
~~a_3 \equiv \beta_{112},~~~
a_4 \equiv \beta_{332}, \\
a_5 &\equiv& \beta_{113},~~~a_6 \equiv \beta_{223},~~~a_7 \equiv \beta_{123}, \\
A &\equiv& \beta_{111},~~~B \equiv \beta_{222},~~~C \equiv \beta_{333}.
\eeqa
We further introduce new variables $r_i$ and $p_i$ which are related
with the first terms  proportional to $\beta \beta$ and
the third terms  proportional to $\alpha$ in equation (\ref{c3}), respectively.
They are defined as follows.
\beqa
r_1&\equiv& 2 h_3 (A a_1-a_3 ^2) + 2 h_2 (A a_2-a_5 ^2)+ 2 h_1 ( a_1 a_2 -a_7
^2),\\
r_2&\equiv& 2 h_3 (B a_3-a_1 ^2) + 2 h_2 (B a_4-a_6 ^2)+ 2 h_2 ( a_3 a_4 -a_7
^2),\\
r_3&\equiv& 2 h_1 (C a_6-a_4 ^2) + 2 h_2 (C a_5-a_2 ^2)+ 2 h_3 ( a_5 a_6 -a_7
^2),\\
p_1&\equiv& -\frac{3}{2} s_1^2(24\lambda_1 g_1+4 \lambda_2 g_2+4 \lambda_3 g_3+
4 \lambda_1 g_2+4 \lambda_1 g_3) \nonumber\\
& &+\frac{6}{5}s_1^2(\lambda_1 +\lambda_2 +  \lambda_3)
(3g_1 +g_2 +g_3) ,\\
p_2&\equiv&-\frac{3}{2}s_1^2(24\lambda_2 g_2+4 \lambda_1 g_1+4 \lambda_3 g_3+
4 \lambda_2 g_1+4 \lambda_2 g_3) \nonumber\\
& &+\frac{6}{5}s_1^2(\lambda_1 + \lambda_2 +  \lambda_3)
(g_1 +3g_2 +g_3),\\
p_3&\equiv&-\frac{3}{2}s_1^2(24\lambda_3 g_3+4 \lambda_2 g_2+4 \lambda_1 g_1+
4 \lambda_3 g_2+4 \lambda_3 g_1) \nonumber\\ 
& & +\frac{6}{5}s_1^2(\lambda_1 + \lambda_2  + \lambda_3)
(g_1 +g_2 +3g_3).
\eeqa
Then neglecting odd terms of $\omega$, which do not contribute to the
average  and using  $O(3)$ symmetry further, we can replace
 $f_{,11} f_{,22}  f_{,33}$ by
\beq
f_{,11} f_{,22}  f_{,33} \Rightarrow r_1  r_2 r_3 +3 (r_1 +p_1)  g_{ij} 
\omega_{ij22}
g_{kl} w_{kl33} +3 p_1 r_2 r_3 + 3  r_1 p_2 p_3 +p_1 p_2 p_3. \lab{c12}
\eeq
In the same manner,  $  f_{,11} f_{,23}^2 $ and $ f_{,12} f_{,32} f_{,13}$
are written in  simpler forms,
\beqa
 f_{,11} f_{,23}^2  &\Rightarrow& (r_1 +p_1) g_{ij} \omega_{ij23} g_{kl} 
\omega_{kl23}
+p_3 r_{12} r_{12}, \\
f_{,12} f_{,32} f_{,13} &\Rightarrow& r_{12}r_{23}r_{13},
\eeqa
where $r_{12}$, $r_{23}$ and $r_{13}$ are  defined 
as
\beqa
r_{12}&\equiv& h_1 (a_1 x_4-a_2 B-2a_6 a_7) +  h_2 (Aa_4+a_2a_3-2a_5a_7 )+  h_3 (AB -a_1a_3 ),\\
r_{23}&\equiv&h_1(BC- a_4a_6 ) +  h_2 (a_4a_5+a_3C-2a_2a_7)+h_3 (a_3a_6+Ba_5 -2a_7a_1),\\
r_{13}&\equiv&h_1(Ca_1+a_6a_2-2a_7a_4  ) +  h_2 (AC-a_2a_5)+h_3 (Aa_6+a_1a_5-2 a_3a_7).
\eeqa

We perform $\omega$- and $\beta$-integral in order.
 In equation (\ref{c12}) only the second term contains $\omega$ which is 
to be replaced by the corresponding variance (eq.\ [\ref{omegacorr}]). 
After some tedious algebra we obtain the following  result.
\beq
\llangle  g_{ij} \omega_{ij22} g_{kl} \omega_{kl33} \rrangle _\omega
 \Rightarrow
-{\frac{144 s_1 ^4}{5}}(46 g_1 ^2 + 184 g_1   g_2+184 g_1   g_3+92 g_2 g_3+
138 g_2 ^2 +138 g_3 ^2).
\eeq
Here we have dropped the terms proportional to $s_2 ^2$, since this cancels 
 with the similar term in (\ref{c12}) (see  MY).

We then move onto $\beta$-integral which is of the form
\beq
\int  \frac{d^{10}\beta}{(2\pi)^5\sqrt{\det\calM}}
\exp\lmk-\frac{1}{2}\beta\calM^{-1}\beta\rmk 
\dirac(f_{,1})\dirac(f_{,2})\dirac(f_{,3})\times(\rm polynomial~ of~ \beta).
\label{betaintegral}
\eeq
Here $\calM^{-1}$ is the inverse of the correlation matrix of $\beta$
whose components are given as follows.
\[336s_1^2{\cal M}^{-1}= \bordermatrix{
      &  A &   a_1  & a_2 &  B & a_3 & a_4 & C  & a_5 & a_6 & a_7  \cr
    A & 2/3 & -1/2  &  -1/2  & 0  &0&0&0&0&0& 0   \cr        
  a_1& -1/2 & 3     & -1/2&0  &0&0&0&0&0& 0  \cr
a_2& -1/2&-1/2&3&0  &0&0&0&0&0& 0  \cr
 B&0&0&0& 2/3 & -1/2  &  -1/2  & 0&0&0&0\cr
a_3&0&0&0& -1/2 & 3     & -1/2&0  &0&0&0\cr
 a_4&0&0&0& -1/2 & -1/2     & 3&0  &0&0&0\cr
 C&0&0&0&0&0&0& 2/3 & -1/2     & -1/2&0 \cr
a_5&0&0&0&0&0&0& -1/2 & 3     & -1/2&0\cr
a_6&0&0&0&0&0&0& -1/2&-1/2&3&0\cr
a_7&0&0&0&0&0&0&0&0&0&7 \cr
} \]
The factor $336$ arises due to our normalization of the variance 
of $\beta$. 
First we use three delta functions $\dirac (f_{,i})$ and replace
$A,~B$, and $C$ by
\beqa
A&=&-(g_2 a_1 +g_3 a_2)/g_1, \nonumber \\ 
B&=&-(g_1 a_3 +g_3 a_4)/g_2, \\ 
C&=&-(g_1 a_5 +g_2 a_6)/g_3, \nonumber 
\eeqa
respectively.  Then the remaining $\beta$-integral is seven
dimensional and we find
\beq
(\ref{betaintegral})=
\frac{1}{g_1g_2g_3}\frac{\sqrt{\det M}}{(2\pi)^{\frac{3}{2}}\sqrt{\det\calM}}
\int \frac{d^{7}a_k}{(2\pi)^\frac{7}{2}\sqrt{\det M}}
\exp\lmk-\frac{1}{2}a_i M^{-1}_{ij}a_j\rmk 
\times({\rm polynomial~ of~} a_k).
\eeq
Here $M_{ij}$ is a correlation matrix of $a_i$ obtained from $\calM$.
  For example, for $i,j=1,2$ we find
\beqa
\llangle a_1 a_1  \rrangle&=&\frac{192s_1^2(9g_1^2+3 g_1 g_3+2g_3^2)}{15
  g_1^2+6g_1 g_2+6g_1 g_3+ 3g_2^2 +2 g_2 g_3+3 g_3^2}\\
\llangle a_1 a_2  \rrangle&=&\llangle a_2 a_1  \rrangle=\frac{96s_1^2(3 g_1^2-g_1 g_2-3g_1 g_3
-4g_2 g_3)}{15 g_1^2+6g_1 g_2+6g_1 g_3+3 g_2^2 +2 g_2 g_3+3 g_3^2}\\
\llangle a_2 a_2  \rrangle &=&\frac{192s_1^2(9g_1^2+3 g_1 g_2+2g_2^2)}{15 g_1^2+6g_1 g_2+6
g_1 g_3+ 3g_2^2 +2 g_2 g_3+3 g_3^2}
\eeqa
The other components are obtained similarly except for
$M_{77}=48$.
As pointed out before this matrix is  constituted from 
$2\times 2  \oplus2\times 2  \oplus
2\times 2  \oplus1\times1$ blocks.

Finally the remaining Gaussian integral over $a_k$ is done by using
the following relations of averages.
\beqa
\llangle a_i a_j a_k a_l \rrangle &=& M^{}_{ij} M^{}_{kl}+
M^{}_{ik}M^{}_{jl}+M^{}_{il}M^{}_{jk} \\
\llangle a_i a_j a_k a_l a_m a_n \rrangle &=& 
 M_{ij} M_{kl}M_{mn}+ \ldots +M_{in} M_{jk}M_{lm},~~~(\rm 15~terms).
\eeqa

Thus we finish $\omega$ and $\beta$ integration in equation (\ref{c12}).
Other terms in equation (\ref{c11}) can be dealt with
 in the same manner. Then we obtain the
final form
\beqa
\llangle N_{st}  \rrangle_{\beta\omega} &=&-(\lambda_1 \lambda_2+
 \lambda_2  \lambda_3+ \lambda_3 \lambda_1 -2D \lambda_1 \lambda_2
\lambda_3) \lab{c43}\\
&\times& \sqrt{\pi^{-5}\cdot 2^{-75} \cdot 3^{-15}\cdot 5^{-1}  \cdot
  7^{-3}} \\
&\times& \exp \left[ - \frac{1}{120} \left\{ -3 \left( \lambda_1+
\lambda_2+\lambda_3 \right)^2  +\frac{15}{2}\left(\lambda_1 \lambda_2 +\lambda_2 \lambda_3
+\lambda_3 \lambda_1\right)\right\}\right]  \nonumber \\
&\times&|(\lambda_1 - \lambda_2)(\lambda_2 - \lambda_3)
(\lambda_3 - \lambda_1)| \lab{c45} \\
&\times&(g_1 g_2 g_3)^{-1}  \dirac (f-c) \sqrt{ {\rm det}M}\lab{c47} \\
&\times&\biggl\{\llangle r_1r_2r_3 \rrangle_{\beta}+ 3\lmk\llangle r_1\rrangle 
_{\beta} +p_1\rmk \lmk - \frac{144}{5} s_1 ^4 \lmk 46 g_1 ^2+ \ldots+138 g_3 ^2
 \rmk \rmk
+3 p_1 \llangle r_2r_3 \rrangle_{\beta} \nonumber \\
& &+3 \llangle r_1 \rrangle_{\beta} p_2p_3+p_1p_2p_3  \lab{c48} \\
& &-3\lmk \llangle r_1 \rrangle_{\beta} +p_1\rmk s_1^4 \lmk -288 g_1^2+ 
+\ldots - 2592g_3^2 \rmk -3 p_3 \llangle r_{12} r_{12} \rrangle_{\beta} \lab{c50} \\
& & +2\llangle r_{12} r_{23} r_{31}  \rrangle_{\beta} \biggl\} \lab{c51}
\eeqa
Here the factor (\ref{c43}) 
comes from $f_{,D}$ term, and the factor (\ref{c45})  from the 
PDF of $\lambda$ (Doroshkevich 1970).
the terms (\ref{c48})
through (\ref{c51}) arise from
${\rm det}f_{,ij}$.
In the above $\llangle N_{st}  \rrangle_{\beta\omega}$, we
have already included the effect of (Euler) angular integral.
So  $\int_0^{D_1}\int\int\int dD d^3\lambda \llangle N_{st}
\rrangle_{\beta\omega}$  corresponds to the evolution of the genus number 
density.

\section{PERTURBATIVE EVALUATION OF THE GENUS CURVE}

In this appendix, we present  perturbative evaluation
of the genus curve for comparison with our result. First we recall
that  Matsubara (1994) obtained correction to the Gaussian formula 
(eq.\ [\ref{lin}]) using the multi-dimensional Edgeworth expansion 
in the second-order Eulerian perturbation theory.  The final result is 
\begin{equation}
   G_{\rm 2nd}(\nu) = - \frac{1}{4\pi^2}
   \left(\frac{\sigma_1^2}{3\sigma^2}\right)^{\frac{3}{2}}
   e^{-\frac{\nu^2}{2}}
   \left[ H_2(\nu)
          + \sigma \left( \frac{S}{6} H_5(\nu)
                          + \frac{3T}{2} H_3(\nu)
                          + 3U H_1(\nu)\right)
          + {\cal O}(\sigma^2)
   \right] .
   \label{eq:second}
\end{equation}
In the above expression, $H_n(\nu) \equiv (-)^n e^{\nu^2/2} (d/d\nu)^n
e^{-\nu^2/2}$ is the $n$-th order Hermite polynomial, and $S$, $T$,
and $U$ are defined as
\begin{eqnarray}
   && S = \frac{1}{\sigma^4} \langle\delta^3\rangle,
   \nonumber \\
   && T = - \frac{1}{2\sigma_1^2\sigma^2}
            \langle \delta^2 \nabla^2 \delta \rangle,
   \label{eq6} \\
   && U = - \frac{3}{4\sigma_1^4}
            \langle \nabla\delta\cdot\nabla\delta
                    \nabla^2 \delta \rangle ,
   \nonumber
\end{eqnarray}
respectively, which we call generalized skewness. They come from the
three-point correlation function that can be evaluated also by
second-order perturbation theory (Matsubara 1994).  The solution of
the Eulerian second order perturbation theory of gravitational
instability in Fourier space is,
\begin{equation}
  \widetilde{\delta}^{(2)}(\bfk,t) =
   D^2
   \int \frac{d^3p_1}{(2\pi)^3} \frac{d^3p_2}{(2\pi)^3}
   (2\pi)^3 \dirac^3(\bfp_1 + \bfp_2 - \bfk)
   P_2(\bfp_1,\bfp_2,t)
   \widetilde{\epsilon}(\bfp_1)
   \widetilde{\epsilon}(\bfp_2),
   \nonumber\\
   \label{eq2-55}
\end{equation}
where 
\begin{equation}
   P_2(\bfp_1,\bfp_2,t) =
   \frac12 \left[
   (1 + K) +
   \frac{\bfp_1\cdot\bfp_2}{p_1 p_2}
   \left(
      \frac{p_1}{p_2} + \frac{p_2}{p_1}
   \right) +
   (1 - K)
   \left(
      \frac{\bfp_1 \cdot \bfp_2}{p_1 p_2}
   \right)^2
   \right].
   \label{eq2-57}
\end{equation}
In the above expressions, $K(t)$ is a weak function of $\Omega$ and
$\Lambda$ (Bouchet et al. 1992; Bernardeau et al. 1995; Matsubara
1995). In Einstein-de~Sitter universe, $K = 3/7$.

The parameters $S$, $T$ and $U$ are then evaluated as
\begin{equation}
   S = 2 (2 + K), \qquad
   T = \frac{2}{3} (5 + 2K), \qquad
   U = \frac{2}{5} (3 + 2K).
\end{equation}
The derivation of the above results of $T$ and $U$ is similar to the
derivation of skewness, $S$ (Bouchet et al. 1992; Peebles 1980).

On the other hand, 
Grinstein \& Wise (1987) give an Eulerian representation of the
Zel'dovich approximation. According to their result, the kernel $P_2$
for Zel'dovich approximation is 
\begin{eqnarray}
   && P_2(\bfp_1,\bfp_2,t) =
      \frac{1}{2}\frac{\bfk\cdot\bfp_1}{p_1^{\,2}}
      \frac{\bfk\cdot\bfp_2}{p_2^{\,2}}
   \nonumber \\
   && \qquad\qquad 
   = \frac{1}{2}\left[
         1 + \frac{\bfp_1\cdot\bfp_2}{p_1 p_2}
         \left(
            \frac{p_2}{p_1} + \frac{p_1}{p_2}
         \right) +
         \frac{(\bfp_1\cdot\bfp_2)^2}{p_1^{\,2}p_2^{\,2}}
      \right].
   \label{eq2-58}
\end{eqnarray}
Thus, $K=0$ corresponds to the Zel'dovich approximation in
second-order Eulerian perturbation theory.

The set of skewness parameters $(S, T, U)$ is $(34/7, 82/21, 54/35)$
for exact second-order perturbation theory (of Einstein-de~Sitter
universe) and $(4, 10/3, 6/5)$ for Zel'dovich
approximation. Therefore, equation (\ref{eq:second}) reduces to
\begin{equation}
   G_{\rm 2nd}(\nu) = \frac{1}{4\pi^2}
   \left(\frac{\sigma_1^2}{3\sigma^2}\right)^{\frac{3}{2}}
   e^{-\frac{\nu^2}{2}}
   \left[ 1 - \nu^2
          + \sigma \nu \left( 
              \frac{4}{5} + 
              \frac{47}{21} \nu^2 - 
              \frac{17}{21} \nu^4
            \right)
   \right] ,
   \label{eq:eds2nd}
\end{equation}
for exact second-order perturbation theory of Einstein-de~Sitter
universe, and
\begin{equation}
   G_{\rm Zel, 2nd}(\nu) =  \frac{1}{4\pi^2}
   \left(\frac{\sigma_1^2}{3\sigma^2}\right)^{\frac{3}{2}}
   e^{-\frac{\nu^2}{2}}
   \left[ 1 - \nu^2
          + \sigma \nu \left( 
             \frac{7}{5} +
             \frac{5}{3} \nu^2 -
             \frac{2}{3} \nu^4
            \right)
   \right] .  
   \label{eq:zel2nd}
\end{equation}
for Zel'dovich approximation.


\newpage

\centerline {\bf REFERENCES}
\bigskip

\def\pp{\par\parshape 2 0true cm 16.5truecm 1truecm 15.5truecm\noindent}
\def\apjpap#1;#2;#3;#4; {\pp#1, #2, #3, #4}
\def\apjbook#1;#2;#3;#4; {\pp#1, #2 (#3: #4)}
\def\apjproc#1;#2;#3;#4;#5;#6; {\pp#1, {\ #2}, #3, (#4: #5), #6}
\def\apjppt#1;#2; {\pp#1, #2}

\apjbook Adler,~R.~J. 1981;The Geometry of Random
Fields;Chichester;Wiley;
\apjpap Bardeen,~J.~M., Bond,~J.~R., Kaiser,~N., \& Szalay,~A.~S. 1986;
ApJ;304;15 (BBKS);
\apjpap Bernardeau,~F. \& Kofman,~L. 1995;ApJ;443;479;
\apjpap Bernardeau,~F., Juszkiewicz,~R., Dekel,~A., \& Bouchet,~F.~R.
1995;MNRAS;274;20;
\apjpap Bouchet,~F.~R., Juszkiewicz,~R., Colombi,~S., \& Pellat,~R.
 1992;ApJ;394;L5;
\apjpap Coles,~P., Melott,~A,~L., \& Shandarin,~S,~F. 1993;MNRAS;260;765;
\apjpap Coles,~P. \& Jones,~B. 1991;MNRAS;248;1;
\apjpap Doroshkevich,~A.~G. 1970;Astrophysics;6;320 (transl. from
Astrofizika, 6, 581);
\apjpap Gott,~J.~R., Melott,~A.~L., \& Dickinson,~M. 1986;ApJ;306;341;
\apjpap Gott,~J.~R., Weinberg,~D.~H., \& Melott,~A.~L. 1987;ApJ;319;1;
\apjpap Gott,~J.~R., Miller,~J., Thuan,~T.~X., Schneider,~S.~E.,
Weinberg,~D.~H., Gammie,~C., Polk,~K., Vogeley,~M., Jeffrey,~S.,
Bhavsar,~S.~P., Melott,~A.~L., Giovanelli,~R., Haynes,~M.~P.,
Tully,~R.~B., \& Hamilton,~A.~J.~S. 1989;ApJ;340;625;
\apjpap Grinstein,B. \& Wise,M.B. 1987;ApJ;314;448;
\apjpap Hamilton,~A.~J.~S. 1988;PASP;100;1343;
\apjpap Hamilton,~A.~J.~S., Gott,~J.~R., \& Weinberg,~D. 1986;ApJ;309;1;
\apjpap Juszkiewicz,~R., Weinberg,~D.~H., Amsterdamski,~P.,
Chodorowski,~M., \& Bouchet,~F. 1995;ApJ;442;39;
\apjproc Kofman,~L.~A. 1994;In Proc.\ Primordial Nucleosynthesis and
Evolution of the Early Universe; ed.\ K.\ Sato;Kluwer; Dordrecht;495;
\apjpap Kofman,~L.~A., Bertschinger,~E., Gelb,~M.~J., Nusser,~A., \&
Dekel,~A. 1994;ApJ;420;44;
\apjppt Loveday,~J. 1996; Preprint astro-ph/9605028 ;
\apjpap Matsubara,~T. 1994;ApJ;434;L43;
\apjpap Matsubara,~T. 1995;Prog.~Theor.~Phys.~(Lett.);94;1151;
\apjpap Matsubara,~T. \& Suto,~Y. 1996;ApJ;460;51;
\apjpap Matsubara,~T. \& Yokoyama,~J. 1996;ApJ;463;409 (MY);
\apjpap Melott,~A.~L., Weinberg,~D.~H., \& Gott,~J.~R. 1988;ApJ;328;50;
\apjbook Milnor,~J.~W. 1965;Topology from the Differential Viewpoint; 
University Press of Virginia, ;Charlottesville;
\apjpap Moore,~B., Frenk,~C.~S., Weinberg,~D.~H., Saunders,~W.,
Lawrence,~A., Ellis,~R.~S., Kaiser,~N., Efstathiou,~G., \&
Rowan-Robinson,~M. 1992;MNRAS;256;477;
\apjpap Okun,~B.~L. 1990;J.~Stat.~Phys.;59;523;
\apjpap Park,~C. \& Gott,~J.~R. 1991;ApJ;378;457;
\apjpap Park,~C., Gott,~J.~R., \& da Costa,~L.~N. 1992;ApJ;392;L51;
\apjbook Peebles, P.~J.~E 1980;The Large-Scale Structure of the
Universe;Princeton University Press;Princeton;
\apjpap Rhoads,~J.~E., Gott,~J.~R., \& Postman,~M. 1994;ApJ;421;1;
\apjpap Shandarin,~S.~F. \& Zel'dovich,~Ya.~B. 1989;Rev Mod Phys;61;185;
\apjproc Shectman, ~S., et al., 1995;in Wide-Field Spectroscopy and 
the Distant Universe;eds.\ S.J.\ Maddox \& A.\ Arag\'on-Salamanca;
 World Scientific;Singapore;98  ;
\apjpap Sorkin,~R.~D. 1986; Phys. Rev.;D33;978;
\apjpap Totsuji, H. \& Kihara, T. 1969;PASJ;21;221;
\apjpap Ueda,~H. \& Yokoyama,~J. 1996;MNRAS;280;754;
\apjproc Vettolani,~G., et al. 1995;In Wide-Field Spectroscopy and 
the Distant Universe;eds.\ S.J.\ Maddox \& A.\ Arag\'on-Salamanca;
World Scientific;Singapore; 115;
\apjpap Vogeley,~M.~S., Park.~C., Geller,~M.~J., Huchra,~J.~P., \&
Gott,~J.~R. 1994;ApJ;420;525;
\apjpap Weinberg, D.~H., Gott, J.~R., \& Melott, A.~L. 1987;ApJ;321;2; 
\apjpap Weinberg,~D.~H. \& Cole,~S. 1992;MNRAS;259;652;
\apjpap White, S.~D.~M. 1979;MNRAS;186;145;
\apjpap Zel'dovich, Ya.~B. 1970;A\& A;5;20;

\newpage

\centerline{\bf FIGURE CAPTIONS}

\bigskip
\begin{description}

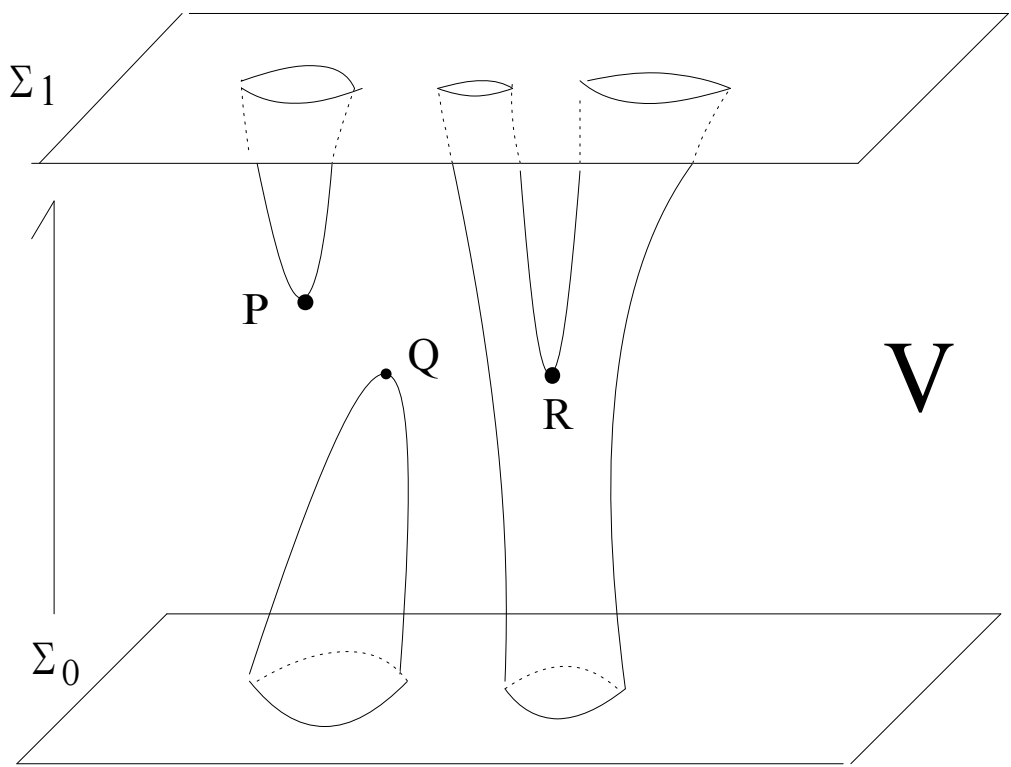
\item[Figure 1]{Schematic explanation of the evolution of the genus
number in which one spatial dimension is omitted.
The genus number on the final surface $\Sigma_1$ is determined by  
the genus number on the initial surface $\Sigma_0$ and 
information of stationary points in the spacetime in between.
At point {\bf P} a new region is created (in this picture the 
signature is $++$). 
At point {\bf Q} a region is annihilated ($--$).
At point {\bf R} a region is split into two regions ($-+$).
In this three dimensional schematic 
picture, merger and split are not distinguished by
signature (both have $-+$),
but in the actual four dimensional case they are different;
merger has $+--$, while split has $++-$.

\label{fig1}}
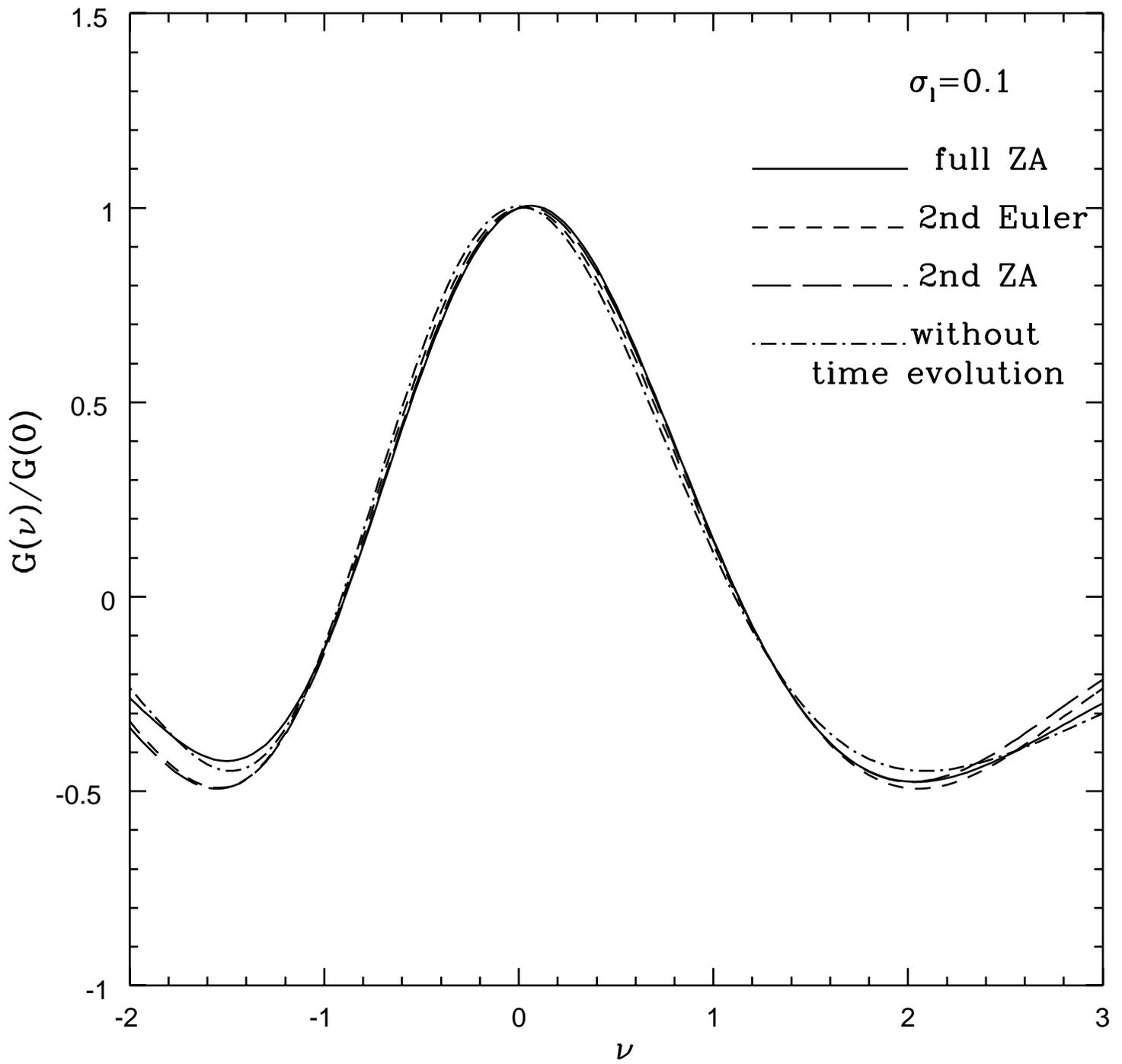
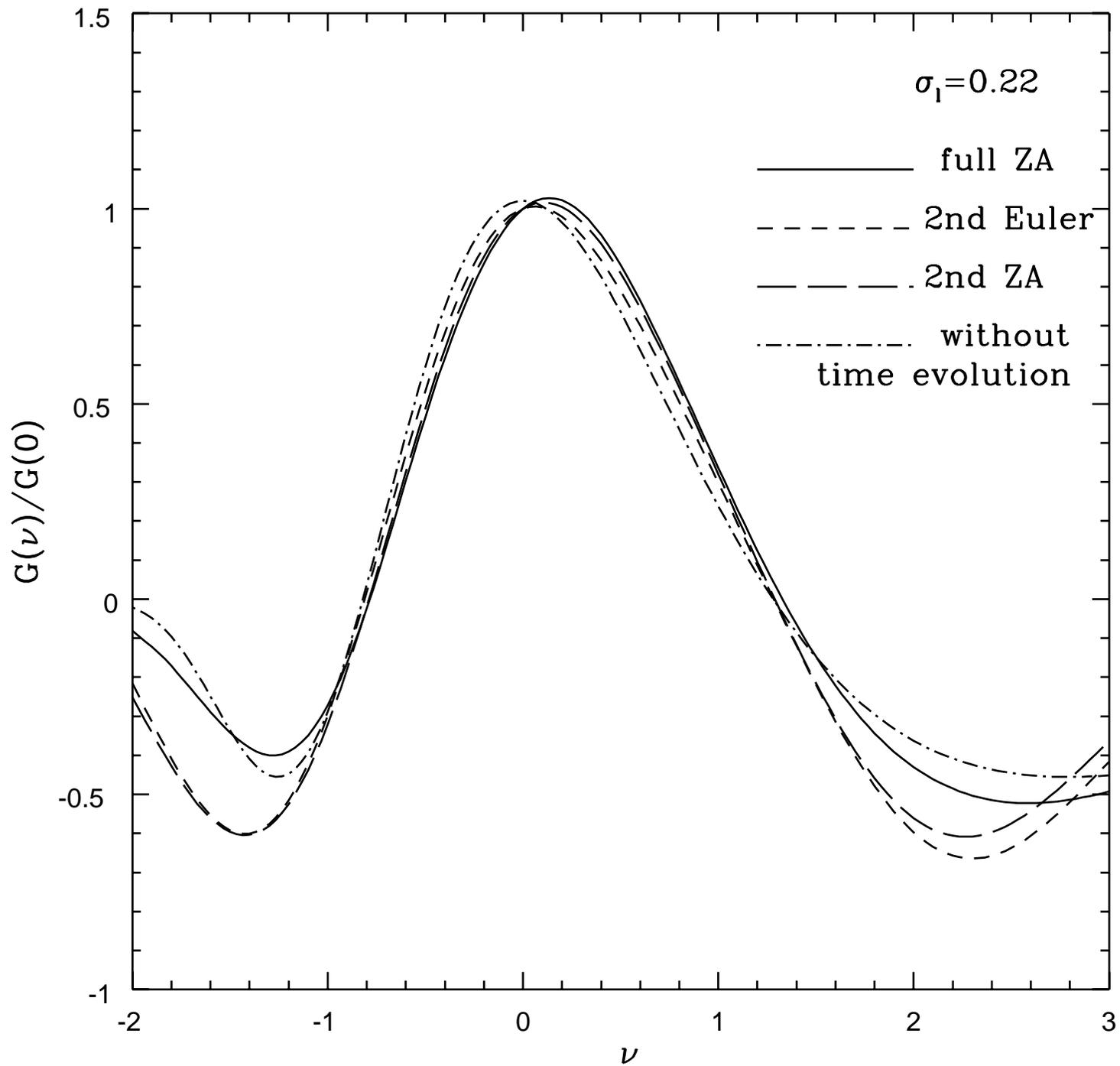
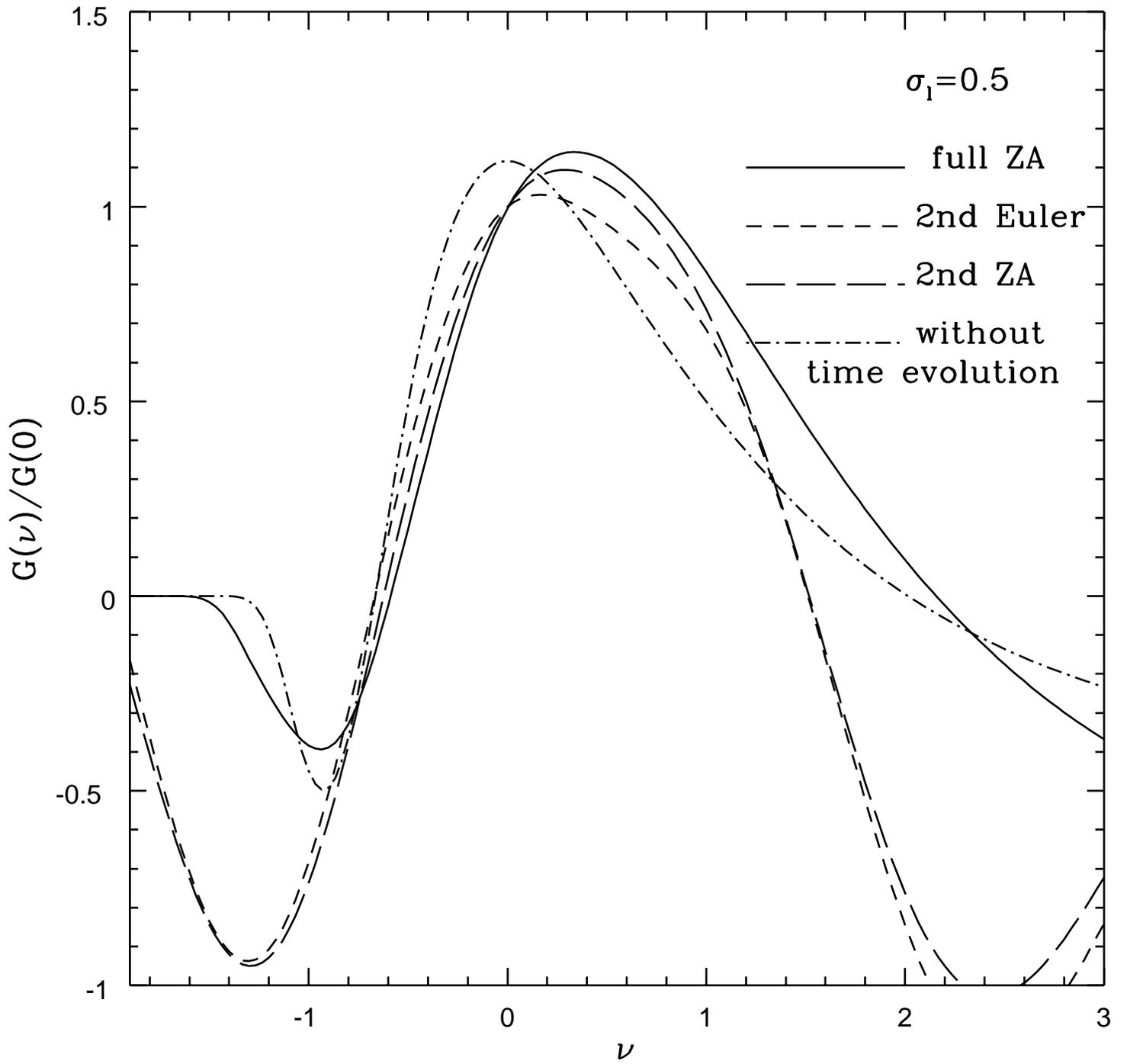
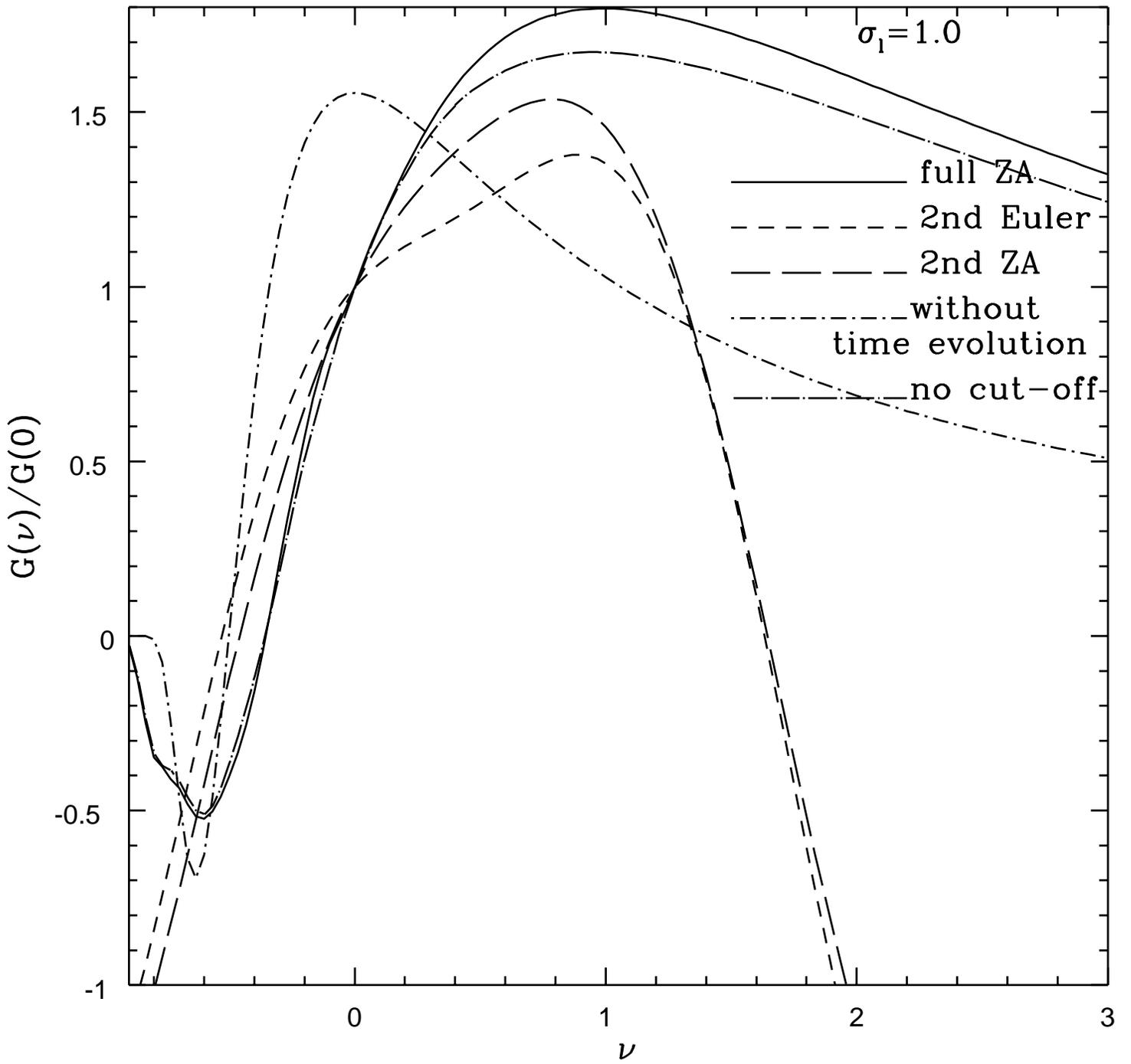
\item[Figure 2]{Normalized genus number density at
$\sigma_l=0.1,~0.22,~0.50,$ and $1.0  $.
Solid line corresponds to our new result of full ZA (eq.\ [\ref{45}]), 
short-dashed line to second-order Eulerian perturbation formula 
$ G_{\rm  2nd}(\nu)$ (eq.\ [\ref{eq:eds2nd}]), long-dashed line to 
second-order perturbation theory for ZA $ G_{\rm Zel,2nd}(\nu)$
(eq.\ [\ref{eq:zel2nd}]), and short-dash-dotted line to the contribution of
the first term of equation (\ref{45}), namely, without the effect of time
evolution of stationary points.  In the last figure with
$\sigma_l=1.0$ long-dash-dotted line represents
 the full ZA result in which $\lambda_i$-integration has been done
up to infinity.

 \label{fig2}}
\item[Figure 3]{
Intuitive explanation for classification of stationary points
in 3+1 spacetime. 
 \label{fig3}}
\end{description}
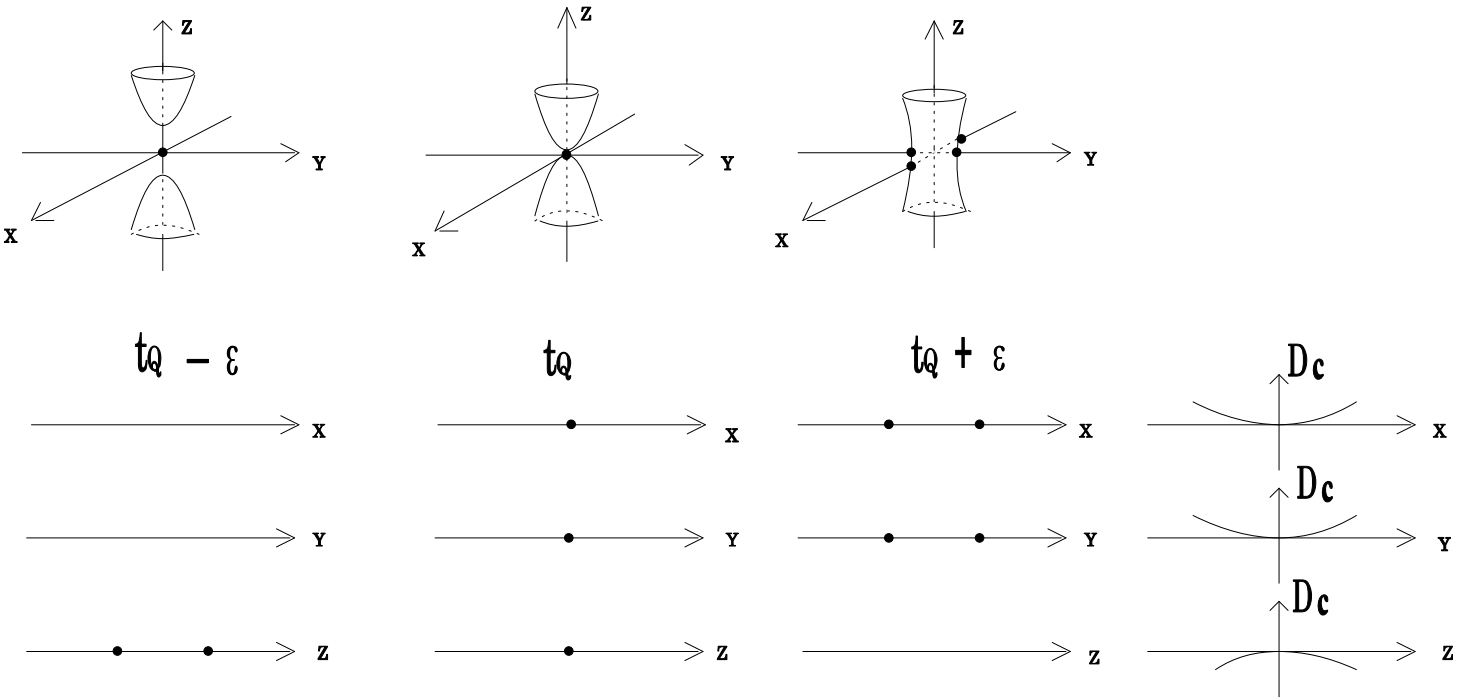
\end{document}